\documentclass[12pt]{iopart}
\input psfig.tex

\usepackage{color}
\usepackage{natbib}
\usepackage{graphicx}
\usepackage{amssymb}
\usepackage{color}
\usepackage{bm}
\definecolor{orange}{rgb}{1,0.5,0}
\definecolor{gray}{gray}{0.50}

\begin{document}

\title[]{Return to axi-symmetry for pipe flows generated after a fractal orifice.}

\author{F C G A Nicolleau\footnote{Corresponding
author: F.Nicolleau@Sheffield.ac.uk}}
\address{SFMG, Department of Mechanical Engineering, The University of Sheffield,
Mappin Street, S1 3JD, Sheffield, UK}

\begin{abstract}
We present experimental results obtained from pipe flows generated by fractal shaped orifices or openings.
We compare different fractal orifices and their efficiencies to re-generate axi-symmetric flows and to return
to the standard flow generated by a
perforated plate or a circular orifice plate.
We consider two families of fractal openings: mono-orifice and complex-orifice and emphasize the differences
between the two fractal families.
For the Reynolds number we use, we found that there is an optimum iteration for the fractal level above which no improvement for practical application such as
flowmetering is to be expected. The main parameters we propose for the characterisation of the fractal orifice are the
connexity parameter, the symmetry angle and the gap to the wall $\delta^*_g$.
The results presented here are among the firsts for flows forced through fractal openings and will serve as reference for future studies and bench marks for numerical applications.
\end{abstract}

\vspace{2pc}
\noindent{\it Keywords}: fractal turbulence, pipe flow, fractal orifice, orifice plate

\maketitle

\section{Introduction}

There have been early attempts at understanding turbulence using fractal formalism
\citep{Sreenivasan-al-1989,Meneveau-Sreeninvasan-1990,Meneveau-1991,Nicolleau-1996},
also by using wavelet analysis \citep{Farge-al93,Nicolleau-Vassilicos99}.

Experiments using fractal-generated turbulent flows have been developed more recently,
these flows are based on geometrically complex, multi-scale shapes, following a fractal pattern.
By contrast to classical turbulence where turbulence is generated using a grid
made of regularly spaced bars or rods, fractal-generated turbulence relies on repeating a pattern at
different scales and positions in the grid.
Such flows seem to point towards different scalings and physics than the
traditional Richardson-Kolmogorov cascade
\citep[for a more complete picture, see e.g.][]{Mazellier-Vassilicos-2010,Valente-Vassilicos-2012}.

In parallel to these new experimental approaches there have been more theoretical attempts
at understanding how particular interactions would generate particular
energy spectra~\citep{Michelitsch-et-al-2009}
%
but we are still far from any complete theory relating fractal and spectral aspects of turbulence.
\\[2ex]
The experimental results presented in this paper are different from other attempts
\citep[e.g.][]{Seoud-Vassilicos-2007}
where the experiments were concerned with homogeneous isotropic turbulence.
Here we concentrate on flow generated in a pipe. The flow is forced through a fractal opening
and we study the effect of that opening on the velocity (mean and rms) profiles.
\\[2ex] Our interest though still
theoretical is also driven by applications.
We are interested in the effect of a complex shape on the generation of turbulent flows in a pipe.
The practical application is to use such shapes as optimal flowmeters or flow mixers.
Studies of complex shaped orifice or lobed jets exist \citep[e.g][]{Ilinca-et-al-2011} here we concentrate on the effect of fractal iterations of a simple pattern.
\\[2ex]
Fractal shapes have been considered as an alternative to the classical circular orifice used
for flowmetering.
Fractal orifices have been shown to decrease pressure drops by as much as 10\%
when compared to the classical circular orifice~\citep{El-Azm-et-al-2010,Nicolleau-et-al-JOT-2011}.
They also improve the measurement quality when used as flowmeter conditioners
\citep{Manshoor-et-al-2011}.
\citep[See e.g.][for another type of non-conventional flowmeters.]{Nowakowski-et-al-AIAA-2011}
A classical orifice flowmeter consists of a circular orifice which is placed in the pipe or duct
to create a pressure drop.
The flow rate is then deduced from this pressure drop. Behind the orifice a substantial
pressure deficit occurs affecting the turbulence cascade process and slowing down the flow
mixing which results in delaying the flow recovery and eventually in a net pressure loss.
\\[2ex] Though the orifice plate as a differential flowmeter has gained an overwhelming popularity
because of its low maintenance cost (no moving part) compared to many other existing flowmeters,
its inherent inefficient flow mixing is a serious disadvange.
Ideally we would like the flow to recover as fast as possible and as much as possible
in order to lower the energy cost associated to the pressure drop necessary for flowmetering.
By contrast, a classic perforated plate will give efficient flow mixing but no pressure drop significant enough for
flowmetering applications.
\\[2ex]
Perhaps a forced (non-Kolmogorov) energy cascade could speed up
the flow recovery by improving the flow mixing. This is where practical interest meets the theoretical research on fractal-generated
turbulence which can provide interesting alternatives for forcing turbulence.
To tackle this problem from the practical point of view, we developed
fractal orifices, thought of as intermediary geometries in between circular orifice and classical perforated plates, and studied their performance as pressure drop generators.

\section{Experimental Set-up}

\begin{figure}[htb]%
\includegraphics[width=13.5cm]{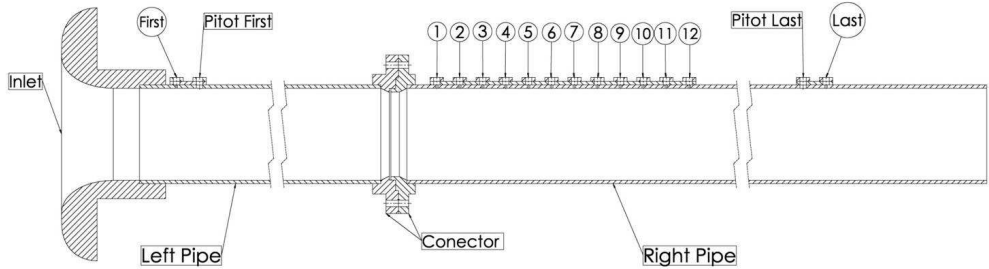}
 \caption{Experimental set up with probe's positions.}
 \label{aiaa-apparatus}
\end{figure}

The wind tunnel and its experimental conditions have been reported in details in~\citep{Nicolleau-et-al-JOT-2011},
we just summarise them here.
The schematic of the wind tunnel is shown in \fref{aiaa-apparatus}.
The 5~mm thick polycarbonate wind-pipe has a length of 4400~mm and an inner diameter $D=140.8$~mm.
For all the plates (Set~1 and Set~2) the experimental initial conditions are exactly the same,
these are summarised in \tref{tab1}.
All plates have the same flow area. The corresponding diameter $d_i$ for the classical circular orifice that is used for comparison is 80~mm.
The inlet velocity is $U_0=5$~\mbox{m s}$^{-1}$.

We can define three Reynolds numbers, the bulk Reynolds number based on the inlet condition:
\begin{equation}
Re_D = {U_0 D \over \nu}
\end{equation}
an orifice Reynolds number based on the orifice diameter and equivalent velocity for the reference circular orifice:
\begin{equation}
Re_d = {\left ( U_0 {D^2 \over d^2} \right ) d \over \nu}
\end{equation}
and a local maximum Reynolds number, $Re_{lM}$, based on the maximum velocity $U_{max}$ and the characteristic length over which it is observed. This Reynolds number is measured at the first station $x/D=0.25$ and depends on the plate (see \tref{tabpara}).
\begin{table}[h]
\caption{\label{tab1} Experimental conditions for the different sets of plates.}
\begin{indented}
\lineup
\item[]\begin{tabular}{@{}*{7}{l}}
\br
\multicolumn{2}{c}{Experimental conditions} \cr
\mr
        Inlet velocity, $U_0$, m~s$^{-1}$ & 5 \cr
        Reynolds number, $Re_D$ & 39800 \cr
        Reynolds number, $Re_d$ & 70000 \cr
        Inlet pressure drop, $\Delta p$, kPa & -11.50 \cr
%
\br
\end{tabular}
\end{indented}
\end{table}
\\[2ex]
There are twelve locations immediately behind the orifice plate (hereafter referred to as \textit{``stations''}) where the measurements are taken as shown in \fref{aiaa-apparatus}.
The first station, Station~1, is $0.25~D$ behind the plate followed by the remaining eleven stations. The distance between two successive stations is $0.25~D$. We conduct hot-wire velocity measurements at these different locations in the pipe and report profiles as functions of the distance from the wall.
\\[2ex]
A motor driven fan was used to suck down the ambient air into the pipe through the bell mouth and leave through a control valve.
The control valve can be regulated to achieve the desired flow velocity in the tunnel.
The inlet velocity is checked before each measurement using hotwire anemometry.

For the sake of comparison, it is important to maintain a single uniform inlet velocity condition that will be kept constant for all the measurements. This condition was well satisfied by the bell-mouth inlet. It ensures that the inlet velocity has a flat profile with a single velocity value $U_0=5$~m~s$^{-1}$.

The profile resolution is $\Delta y = 5$~mm that is $\Delta y^* = \Delta y/D = 0.0625$. This was chosen as an optimum between experimental time constraint and profile resolution. It is also consistent with the total width of the hotwire which is of the order of few mm.

\section{Fractal orifices}

We study two different types of orifice plates, the `mono-orifice' and the `complex-orifice'.
By `mono-orifice' we mean a single connex hole cut on a plate, by `complex-orifice' we mean a series of disconnected holes cut on a plate.
\\[2ex]
A mathematical fractal shape is obtained by iterating a pattern at smaller and smaller scales. Strictly speaking the `fractal object' is obtained in the limit of an infinite number of iterations. We adopt this construction approach for our `fractal' orifices and consider different iterations of the pattern corresponding to different achievements of the fractal geometry.
\\[2ex]
i) The fractal mono-orifices that we use in Set~1 can be thought of as a variation
of the classical (smooth) circular orifice which we consider as our reference.
We modified the perimeter of the circular orifice to add sharpness and irregularity to this smooth circle.
Hence, similarly to the iteration levels of the fractal construction, the edges of the fractal orifice get sharper and sharper at each iteration level.
\\[2ex]
ii) The fractal complex-orifices in Set~2 make reference to the classical perforated plate with same-sized holes uniformly drilled on the whole surface of the plate.
By contrast to the mono-orifice in this case there is no connected flow area.
\\[2ex]
For each set a total of four fractal orifices $(N = 0-3)$ was considered for the present study. All the plates have the same porosity. It is worth noting that when the iteration increases, in order to maintain the porosity the object needs to be scaled down. That is the generator-length in the fractal construction has to be decreased.

\subsection{Fractal mono-orifice, Set 1}

\begin{figure}[h]
\begin{center}
\includegraphics[width=13.5cm]{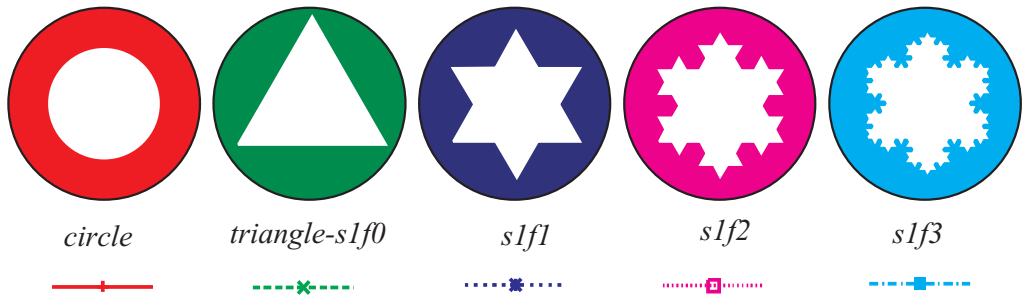}
\end{center}
\caption{Set~1: classical circular orifice plate, fractal orifice plates with four levels of iteration,
\textit{s1f0}, \textit{s1f1}, \textit{s1f2}, \textit{s1f3}.}
\label{Set1}
\end{figure}

The fractal orifices that we used for Set~1 (\fref{Set1}) are based on the von Koch pattern. This fractal pattern is characterized by a fractal dimension of 1.26. The different iterations of fractal orifices for Set~1 are shown and labelled in \fref{Set1}.
All the plates have an equal flow area.
\\
The different plates constituting this first set can be classified as
orifices, i.e. they can be thought of as a hole with a fractal perimeter. So the natural reference for comparison is the classical circular orifice plate which is the first plate in this set, whereas the second plate \textit{s1f0} corresponds to the initial pattern which is just a triangular hole.
All the sets except the triangle are symmetric with respect to both the horizontal and vertical planes.

\subsection{Fractal complex-orifices, Set 2}

Set~2 obeys a different rule, it corresponds to a fractal distribution of holes of different scales. So the natural standard for comparison is a turbulence generated by a series of uniformly distributed identical holes as in the classical perforated plate. The first step which defines the fractal pattern is here just the classical circular orifice.
It is labelled \textit{circle-s2f0} but is nothing but the previous first plate of Set~1.
By contrast to Set~1, Set~2 is more a fractal grid than strictly speaking a fractal orifice.
The plates have equal flow area and porosity, the same as in Set~1.
\noindent
\begin{figure}[h]
\includegraphics[width=13.5cm]{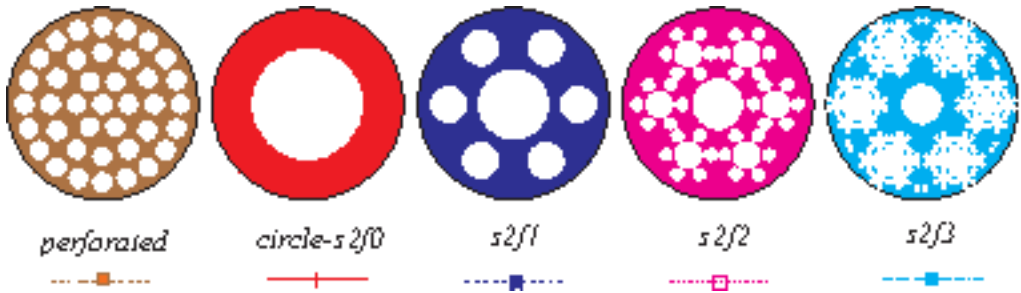}
\caption{\label{Set2}Set~2: classical perforated plate, fractal complex-orifices with four levels of iteration,
\textit{s2f0}, \textit{s2f1}, \textit{s2f2}, \textit{s2f3}.}
\end{figure}

\subsection{Plates classification}

From a practical point of view there are two main parameters to consider: the
Reynolds number and the porosity. We showed in \citep{Nicolleau-et-al-JOT-2011}
that the shape of the orifice can
lead to very different pressure drops and
flow patterns and a fractal approach might be a systematic way to classify those shapes
\footnote{In that respect the classical perforated plate does not introduce a significant pressure drop and is used here only as a benchmark for assessing the
flow return to axi-symmetry.}.
We use the connexity parameter defined in \citep{Nicolleau-et-al-JOT-2011}
indicating when a plate is perforated-like or orifice-like.
It is defined as the inverse of the number of holes or closed areas.
\\[2ex]
\cite{Nicolleau-et-al-JOT-2011} also defined a parameter characterising axi-symmetry, this was measured by the angle of symmetry, that is the smallest angle of rotation leaving the orifice geometry invariant.
These non-dimensional parameters are reported \tref{tabpara}.
\begin{table}[h]
\caption{\label{tabpara} Non-dimensional characteristics for the different plates.}
\lineup
\begin{tabular}{@{}*{2}{l}}
\br
Set &
\begin{tabular}{|p{2cm}|p{1.5cm}|p{2.5cm}|p{2cm}|p{1cm}|p{1cm}|p{1cm}}
plate & porosity & symetry & connexity & scale & $Re_{lM}$ & $\delta_g^*$
\\
      & & angle & parameter & ratio & & gap
\end{tabular}
\cr
\mr
1 &
\begin{tabular}{|p{2cm}|p{1.5cm}|p{2.5cm}|p{2cm}|p{1cm}|p{1cm}|p{1cm}}
circular & 0.336 & 0 & 1 & 1 & 31\,000 & 0.21 
\\
triangle & 0.336 & 2.094 ($2\pi/3$) & 1 & 1 & 29\,700 & 0.05 
\\
s1f1 & 0.336 & 1.047 ($\pi/3$) & 1 & 0.333 & 29\,100 & 0.11 
\\
s1f2 & 0.336 & 1.047 ($\pi/3$) & 1 & 0.111 & 29\,360 & 0.13 
\\
s1f3 & 0.336 & 1.047 ($\pi/3$) & 1 & 0.037 & 27\,500 & 0.14 
\\
\end{tabular}
\cr
\cr
2 &
\begin{tabular}{|p{2cm}|p{1.5cm}|p{2.5cm}|p{2cm}|p{1cm}|p{1cm}|p{1cm}}
circle & 0.336 & 0 & 1 & 1 & 31\,000 & 0.21 
\\
s2f1 & 0.336 & 1.047 ($\pi/3$) & 0.1429 & 0.541 & 14\,000 & 0.06 
\\
s2f2 & 0.336 & 1.047 ($\pi/3$) & 0.0233 & 0.257 & 11\,500 & 0.05 
\\
s2f3 & 0.336 & 1.047 ($\pi/3$) & 0.0037 & 0.111 & 8\,900 & 0.04 
\\
perforated & 0.336 & 0.033 ($2\pi/19$) & 0.026 & 1 & 7\,600 & 0.04 
\end{tabular}
\cr
\br
\end{tabular}
\end{table}
\\[2ex]
$Re_{lM}$ is more or less constant for the mono-orifices. This is to be expected as they are basically a variation of the
circular orifice: one hole size and one maximum velocity observed around the pipe axis. For the complex-orifices, $Re_{lM}$ varies from large values (as a circular orifice) to small values (as a perforated plate). This is the classical trend for Set~2 which combines features from orifice and perforated plates.

\subsection{Velocity profiles}

The velocity measurements were conducted for the two sets of orifices using {constant temperature} hot-wire anemometry
which allows a detailed analysis of the effect of the fractal pattern on the flow experiencing the fractal orifice. {The hot-wire probe that we used is the 55P16 uni-directional single sensor probe from Dantec Dynamics and the constant temperature anemometer type 54T30.
The probe sensing wire is made of platinum-plated tungsten with a diameter of 5~$\mu$m and 1.25~mm long. That is a sensing length to diameter ratio of 250.
It can measure velocities from 3 to 50 m~s$^{-1}$ at a maximum frequency of 10~kHz.
The probe was calibrated before each measurement campaign.
\\[2ex]
We use a single hotwire, the wire is orthogonal to the mean flow direction (pipe axis). Thus we measure a velocity amplitude $U$ that contains the streamwise and the vertical component of the velocity only.}
The measurements were taken on the centerline and the different statistics are reported as functions of the non-dimensional distance $x/D$ from the plate location.
As a reference the velocity was also measured when there was no plate, this corresponds to the data referred to as `np' (i.e. no plate) in the different plots.
\\[2ex]
The flow in our experiment is not homogeneous, so all the statistics presented are averages in time. We checked over long time samples that our signal is steady.
In practice, the time series sampling frequency was $10^4$~Hz. Statistics were performed over more than $10^6$ points that is for more than 100~s which was large enough to obtain converged statistics. {This is more than $10^4$ the integral time scales
as shown in \citep{Nicolleau-et-al-JOT-2011}.}
Velocity series are recorded using classical hotwire anemometry.
The hotwire is positioned orthogonal to the streamwise direction.
So that the recorded velocity takes into account flow fluctuations along two directions and is given by
\begin{equation}
U_{hwa}(t) = \sqrt{U_{x}^2(t) + U_{y}^2(t)}
\end{equation}
where $x$ is the streamwise axis ad $y$ the ver\let\ifpdf\relaxtical axis.
$U_{hwa}$ is recorded as a time series.
This velocity as a time series can be decomposed into its mean and fluctuation parts:
\begin{equation}
U_{hwa} = \overline{U}_{hwa} + u
\label{turbdecompo}
\end{equation}
In this paper we present results for the mean value of
$U_{hwa}$ that we note just ${U}$:
\begin{equation}
{U} = \overline{U}_{hwa}
\end{equation}
in section~\ref{velmean1prof}.
We also present results for the rms of the fluctuation:
\begin{equation}
u' = \sqrt{\overline{u^2}}
\end{equation}
in section~\ref{rmsprof}.
In most plots we use no-dimensional quantities: the velocities are normalised by the inlet velocity $U/U_0$ and the distances are
normalised by the pipe diameter, $x^* = x/D$ and $y^* = y/D$.

\section{Mean velocity profiles}
\label{velmean1prof}

\subsection{Velocity profiles for Set~1}

%
\begin{figure}[ht!]
\includegraphics[width=14.5cm]{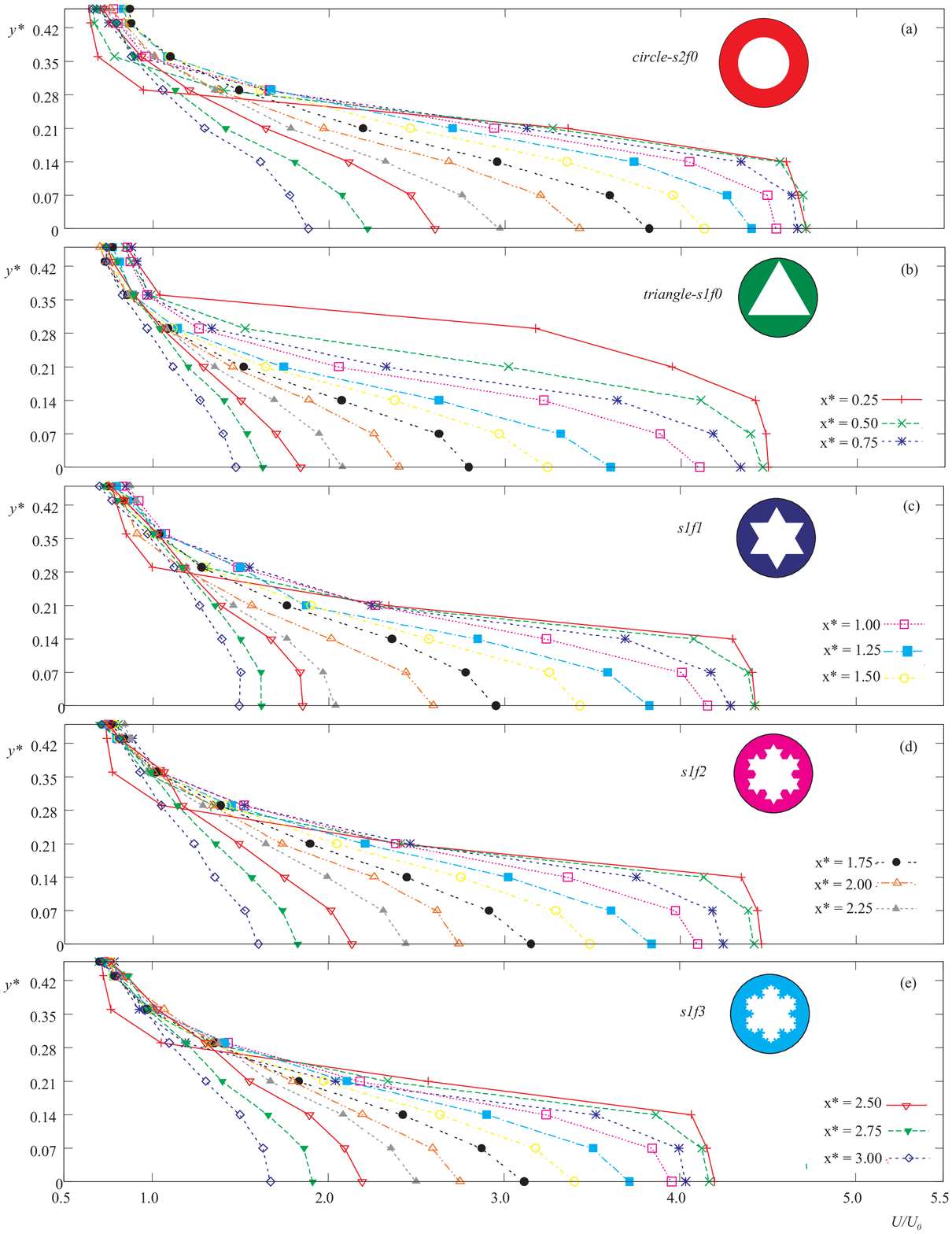}
\caption{Set~1: Mean velocity profile ${U}/U_0$ for different distance $x^*$:
\textcolor{red}{$+$ 0.25}, \textcolor{green}{$\times$ 0.5}, \textcolor{blue}{$\ast$ 0.75},
\textcolor{magenta}{$\square$ 1}, \textcolor{blue}{$\blacksquare$ 1.25}, \textcolor{yellow}{$\opencircle$ 1.5},
$\fullcircle$ 1.75, \textcolor{orange}{$\opentriangle$ 2},
\textcolor{gray}{$\blacktriangle$ 2.25}, \textcolor{red}{$\opentriangledown$ 2.5},
\textcolor{green}{$\blacktriangledown$ 2.75} and \textcolor{blue}{$\diamond$ 3}.}
\label{s1profilebyinlet}
\end{figure}
\Fref{s1profilebyinlet}, shows the mean velocity profiles, $U$, normalized by the inlet velocity, $U_0$ (horizontal axis in the figure), as a function of the distance from the wall $y* = y/D$ (vertical axis in the figure); $y^*=0$ at the axis. The vertical increment is
5~mm, that is $\Delta y^* = 0.07$.
The different curves correspond to the different
downstream locations $x^*$ behind the plates, namely:
$+$ 0.25, $\times$ 0.5, $\ast$ 0.75, $\square$ 1, $\blacksquare$ 1.25, $\opencircle$ 1.5,
$\fullcircle$ 1.75, $\opentriangle$ 2, $\blacktriangle$ 2.25, $\opentriangledown$ 2.5, $\blacktriangledown$ 2.75 and $\diamond$ 3.
\Fref{s1profilebyinlet}a, b, c, d and e correspond respectively to the different orifices:
\textit{circle}, \textit{triangle}, \textit{s1f1}, \textit{s1f2} and \textit{s1f3}.
\\[2ex]
The first thing to notice when looking at the different profiles is that
the triangular orifice plate \textit{s1f0} has to be set apart.
The profiles' evolution for all the other plates is similar.
For these plates, at $x^*=0.25$, the magnitude of the velocity near the wall is smaller than at other stations.
This is because near the wall and near the plate a re-circulation zone exists. This re-circulation zone vanishes further away from the plate.

In the case of \textit{s1f0-triangle}, the vertex of the triangular flow area disrupts the re-circulation zone near the wall, we will see more evidence of this latter on.
The asymmetrical flow area of \textit{s1f0} could also be the reason behind its atypical behaviour.

All the plots are on the same scale, so
another noticeable feature is that the velocity decreases from (a) to (e). Hence the higher the fractal iteration level the smoother the perturbation to the flow. This trend however is not true for all distances from the plate and we will discuss this in more details with
\fref{s1cominlet}.

From \fref{s1profilebyinlet} we can also see that the first two profiles $x^*=0.25$ and $x^*=0.5$ behind the \textit{circular} orifice plate,
collapse on each other for $y^*>0.25$. This is the well-known ``vena contracta'' effect for circular orifices.
This effect is also present in the other orifices, except again for the triangular-shaped orifice.
The vena contracta effect is more pronounced in the case of the circular orifice with close profiles up to $x^*=0.75$,
whereas for the fractal orifices it disapears after $x^*=0.5$.
One possible explanation is that the circular orifice can be described as monoscale, meaning that it can be
defined by one scale only, its diameter; whereas the other plates can be described as multiscale in that they are forcing on a range of scales.
The fractal distribution of sharp edges is thought to enhance mixing as observed in the numerical experiment of \cite{Melick-Geurts-2012}.
%
%
\subsection{Measure of self-similarity in profiles for Set~1}

\begin{figure}[ht!]
\includegraphics[width=15.cm]{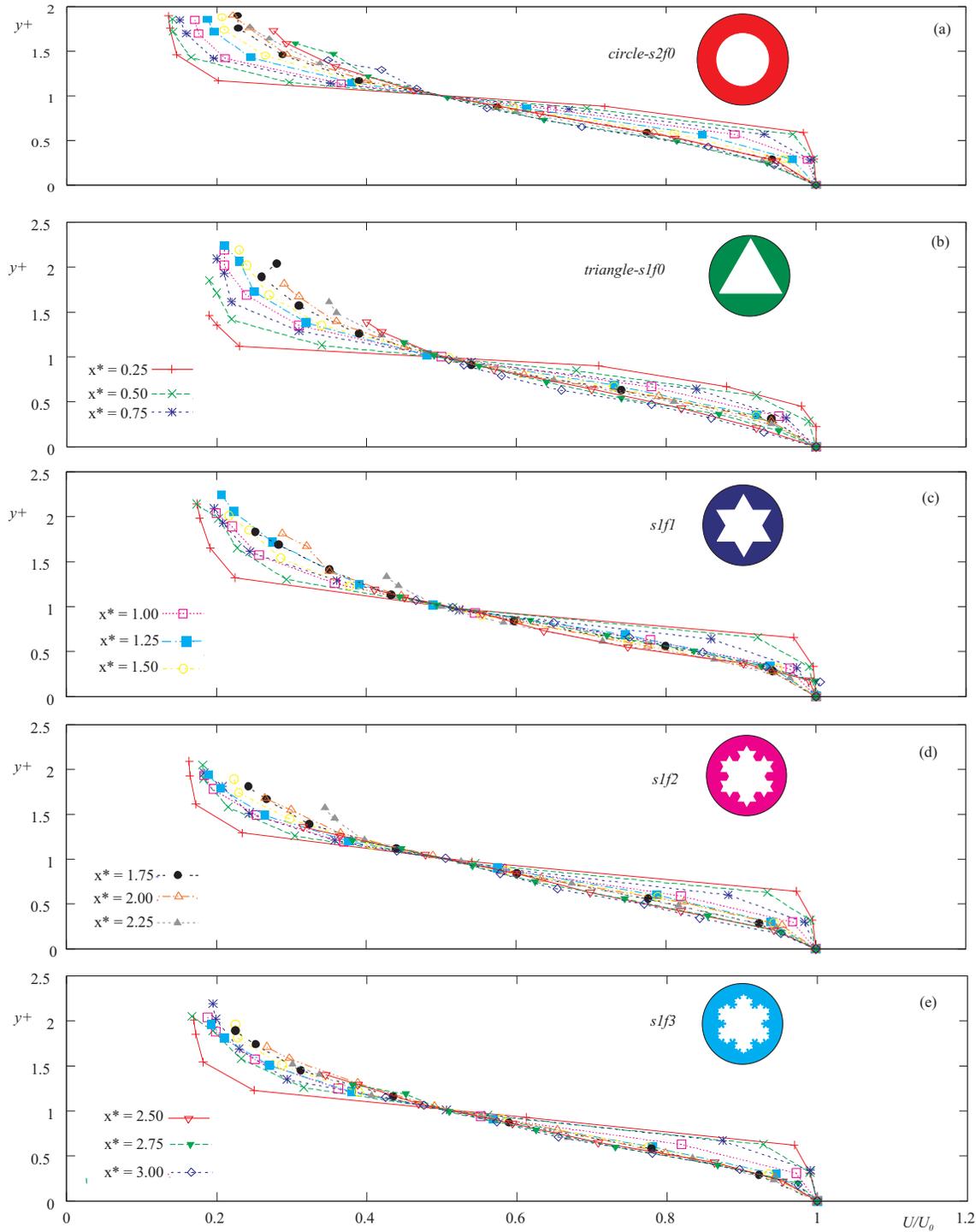}
\caption{Set~1: normalized velocity profiles (${U^*}$, $y*$), for different distance $x^*$:
\textcolor{red}{$+$ 0.25}, \textcolor{green}{$\times$ 0.5}, \textcolor{blue}{$\ast$ 0.75},
\textcolor{magenta}{$\square$ 1}, \textcolor{blue}{$\blacksquare$ 1.25}, \textcolor{yellow}{$\opencircle$ 1.5},
$\fullcircle$ 1.75, \textcolor{orange}{$\opentriangle$ 2},
\textcolor{gray}{$\blacktriangle$ 2.25}, \textcolor{red}{$\opentriangledown$ 2.5},
\textcolor{green}{$\blacktriangledown$ 2.75} and \textcolor{blue}{$\diamond$ 3}.}
\label{s1profilebymax}
\end{figure}

In \fref{s1profilebymax} we use the classical velocity normalisation to look for some trends of self-similarity in the mean velocity profile, on the horizontal axis we report the velocity divided by the axial mean velocity (which is also the maximum velocity):
\begin{equation}
U^* = {{U}(y) \over {U}(0)}
\end{equation}
on the vertical axis we report the distance to the wall normalised by the distance at which the velocity is half that at the centerline
\begin{equation}
y^+ = {y \over y_{1 \over 2}}
\end{equation}
where ${U}(y_{1 \over 2}) = {U}(0)/2$ by definition. So that all normalised profiles go through the fixed point
$(U^*=1/2, y^+=1)$. The spreading of the profile around that fixed point is an indication of how self-similar they are.
\\[2ex]
The notations and symbols are the same as in~\fref{s1profilebyinlet}.
The different curves correspond to the different
downstream locations behind the plates $x^*$, namely:
$+$ 0.25, $\times$ 0.5, $\ast$ 0.75, $\square$ 1, $\blacksquare$ 1.25, $\opencircle$ 1.5,
$\fullcircle$ 1.75, $\opentriangle$ 2, $\blacktriangle$ 2.25, $\opentriangledown$ 2.5, $\blacktriangledown$ 2.75 and $\diamond$ 3.
\Fref{s1profilebyinlet}a, b, c, d and e correspond respectively to
\textit{circle}, \textit{triangle}, \textit{s1f1}, \textit{s1f2} and \textit{s1f3}.
\\[2ex]
The values of $y^+$ are smaller than 2 for the circular orifice, whereas there are values above 2 for the other orifices, so near the wall,
the profiles are less spread for the circular orifice than for the other orifices.
However, far from the wall, the circle and triangle plates show the most spread profiles, whereas for the fractal orifices
\textit{s1f1}, \textit{s1f2} and \textit{s1f3} the profiles tend to a universal shape which is very close to
\begin{equation}
U^* \simeq 2 y^+
\end{equation}
for $0.3 < y^+ < 0.9$.
For the three highest fractal iterations a universal profile is nearly achieved after one diameter. The higher the iteration the better the collapse of the profile for $x^* > 1$.
The upper parts of the profiles ($y^+ > 1$) show more spread. This corresponds to data measured close to the wall which are
less accurate owing to the wall boundary condition: $U \to 0$ when $y^* \to 0.5$, whereas our hotwire limitation is around
3~m~s$^{-1}$.

\subsection{Fractal scaling effects at each station}

It is not easy to see directly the fractal scaling effects on the profile from the previous figures.
%
\begin{figure}[ht]
\includegraphics[angle=0, width=15cm]{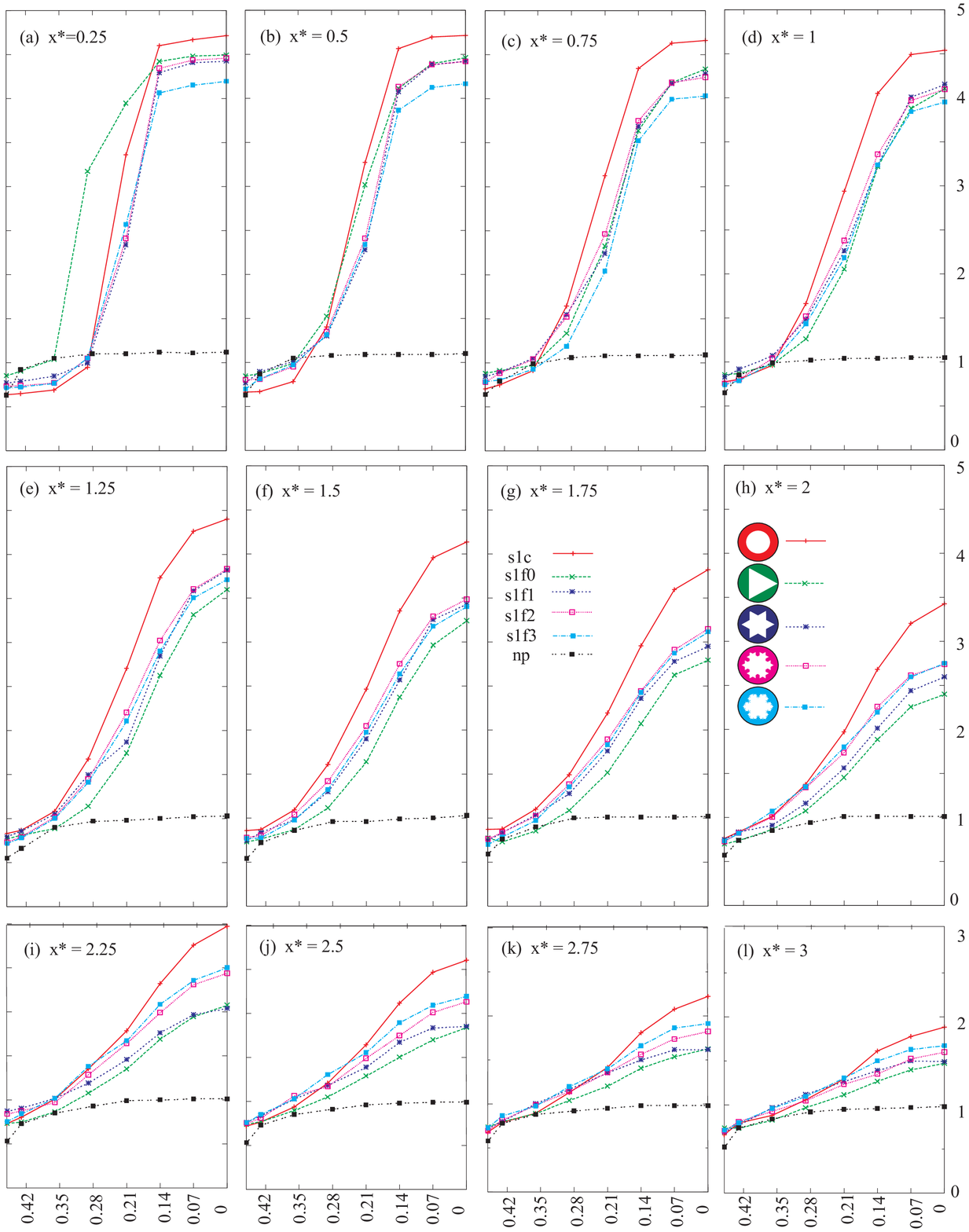}
\caption[]{\label{s1cominlet}
Set~1: Comparison of fractal iteration level effects on velocity profiles ($y^*$, ${U}/U_0$)
for the different stations.}
\end{figure}
To make it clearer in \fref{s1cominlet} we compare the different iteration levels at a given station. The velocity is normalized by the inlet velocity (${U}/U_0$ on the vertical axis) and is plotted as a function of $y^*$.
\Fref{s1cominlet} presents the same profiles as in \fref{s1profilebyinlet} but this time grouped by stations so that we can discuss directly the effect of the fractal iteration level on the profiles. For the sake of comparison the profile measured without any plate has also been included. The different figures (a-l), correspond respectively to
$x^*=0.25$, 0.5, 0.75, 1, 1.25, 1.5, 1.75, 2, 2.25. 2,5. 2.75 and 3. The color coding and line symbol correspond to that in
\fref{Set1}.
\\[2ex]
The figure shows that far from the orifice, $x^*=3$, the circular orifice shows the lowest decrease in velocity and the triangle orifice the
largest decrease. The triangular orifice decays more slowly initially and then has a more rapid decay further downstream.
The fractal-orifices show intermediary behaviour:
\begin{itemize}
\item[i)]
$x^* \le 1$ close to the orifice, the higher the fractal iteration the lower the velocity level,
\item[ii)]
$x^* \ge 2$ far from the orifice, the higher the fractal iteration the higher the velocity level.
\end{itemize}
Further down than $x^*=2.25$ there is little difference between the fractal orifices for $y^*>0.14$.

\subsection{Set~2: Mean velocity profiles}

Figure~\ref{s2profilebyinlet} shows the mean velocity profile for Set~2, ${U}$,
normalized by the inlet velocity, $U_0$,
as a function of the distance from the pipe axis $y^*$ and for the different downstream locations $x^*$.
%
\begin{figure}[ht!]
\includegraphics[width=15cm]{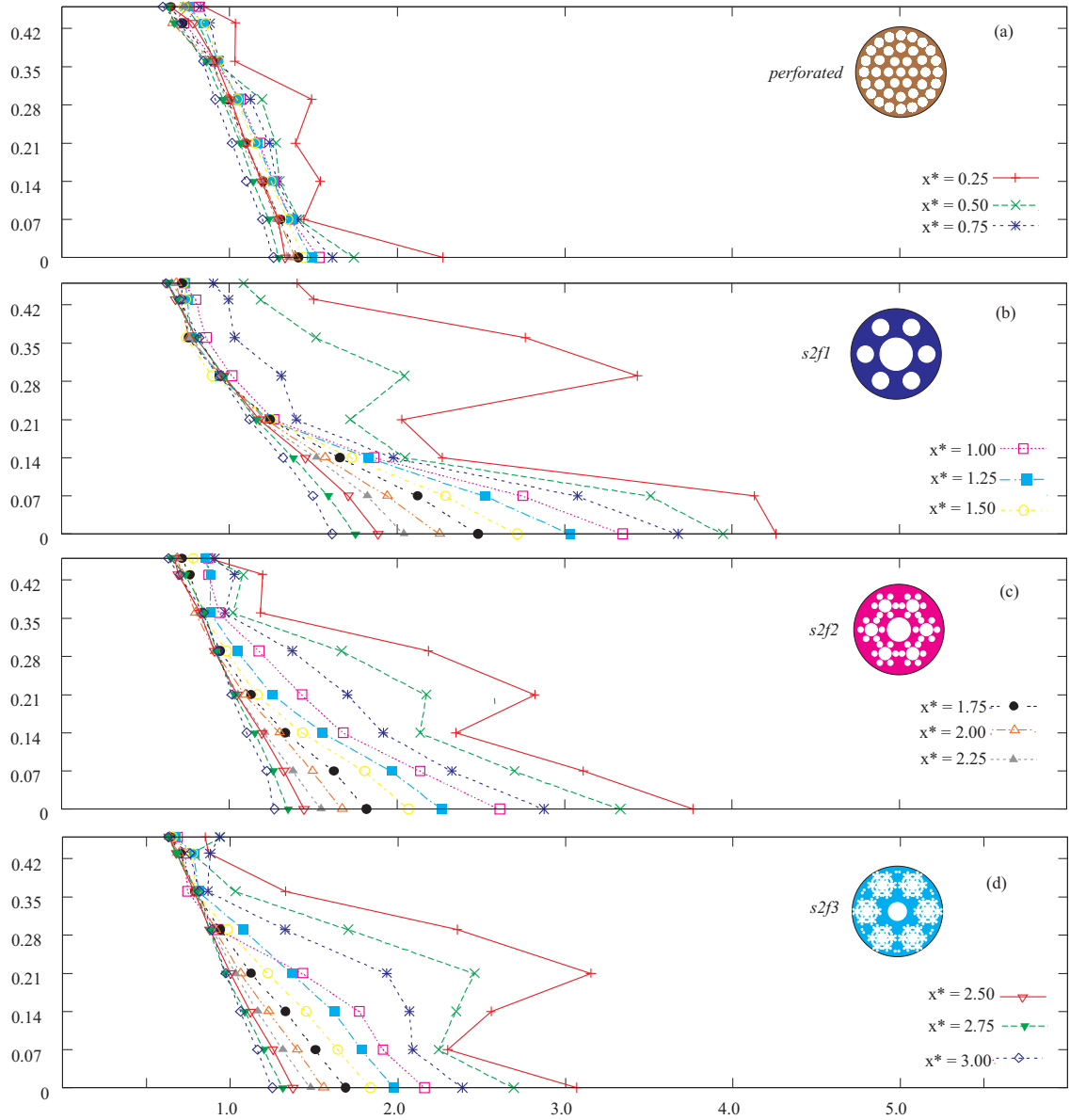}
\caption{Set~2: Velocity profiles ${U}/U_0$ (horizontal axis) as functions of $y^*$ (vertical axis).}
\label{s2profilebyinlet}
\end{figure}

The top panel is the reference classical perforated plate that is often used as a flow conditioner.
The different plots correspond respectively to (a) the classical perforated plate, (b) \textit{s2f1},
(c) \textit{s2f2} and (d) \textit{s2f3}. We omitted the circular orifice in this figure the results for that plate have already been
shown with the results for Set~1 (\fref{s1profilebyinlet}).
\\[2ex]
Near the plate,
the profiles for the fractal complex-orifices show more than one peak as for the perforated plate.
This illustrates the fundamental difference between Set~2 and Set~1. In Set~1 the flow is acted upon from
the orifice's perimeter, whereas Set~2 corresponds to jets fractally distributed in space and scale interacting with each other.
The peaks are smoothed out after $x^*=1$ for all the plates, with no intermediary velocity peak between the wall and the pipe centre.
\\[2ex]
In \fref{s2cominlet} we compare the effect of each complex-orifice at a given station.
The first observation is that the mean velocity near the pipe axis decreases as the fractal iteration level increases.
This can be observed consistently at each station.
\begin{figure}[ht!]
\includegraphics[width=15cm]{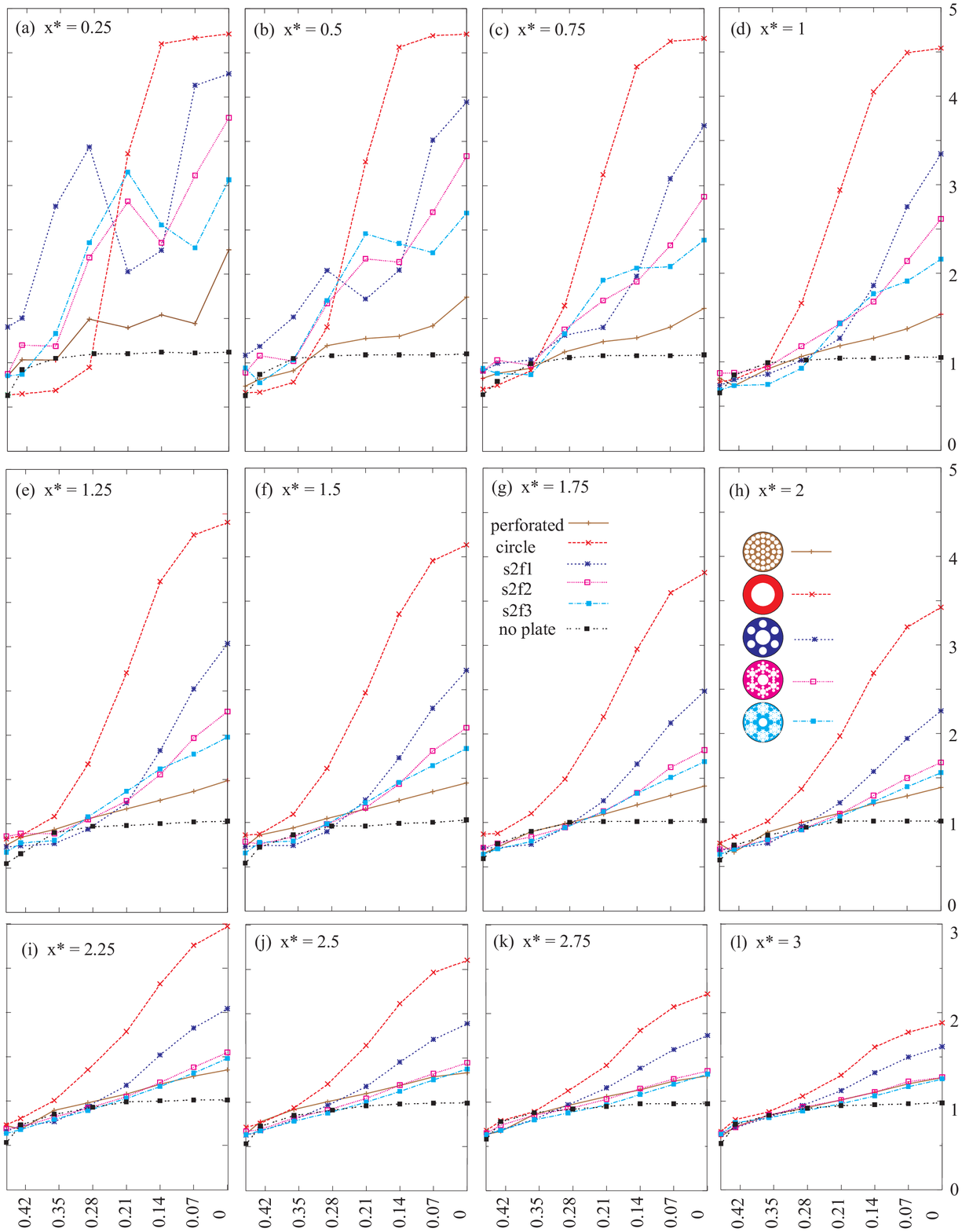}
\caption{Set~2: Comparison of scaling effects on velocity profile normalized by inlet velocity, $U_0$, at each station.}
\label{s2cominlet}
\end{figure}
\\[2ex]
The three fractal complex orifices \textit{s2f1}, \textit{s2f2} and \textit{s2f3} loose their two-peak profile shape at about the same station $x^*=0.75$. \textit{s2f2} and \textit{s2f3} have very close profiles at all stations with peaks at the same $y^*$.

The return to a low smooth profile is faster across the classical perforated plate until $x^*=2$.
Then as the fractal iteration level increases, the velocity profiles get closer to that of the
reference perforated plate.
However, the velocity profiles of the lower fractal iteration level \textit{s2f1} depart significantly from those of
\textit{s2f2}, \textit{s2f3} and the classical perforated plate, even as far downstream as $x^*=3$. \textit{s2f1} keeps a behaviour intermediary between the orifice and the perforated plate.
There is little difference between the profiles of \textit{s2f2} and \textit{s2f3} after $x^*=1.75$ indicating that there is
little to gain in manufacturing plates for higher fractal order than 2. This is consistent with findings in \citep{Nicolleau-et-al-JOT-2011}.
Of course this conclusion is only valid for the Reynolds number we have here. As the Reynolds number increases one can expect more discrepancies between \textit{s2f2} and \textit{s2f3}.
\\[2ex]
To summarise, the fractal complex-orifices \textit{s2f2} and \textit{s2f3} generate high velocity values near the plate, values that are close to the circular orifice values; but further downstream, $x^* > 2$, the velocity decreases rapidly down to the level of the perforated plate. There is a clear effect of the fractal iteration as this is not true for \textit{s2f1}.

\section{Guessed volume flow rates}

As the density is constant in our experiment the mass conservation reduces to the volume flow rate conservation.
That is
\begin{equation}
{\cal V}(x) = \int_0^{1}\int_{0}^{2 \pi} \overline{U}(x,y,\phi) \, y dy d\phi
\label{flowraxir}
\end{equation}
is independent of $x$. Furthermore,
when the flow is axi-symetric the volume flow rate ${\cal V}$ (\ref{flowraxir}) reduces to:
\begin{equation}
{\cal V}(x) = 2 \pi \int_0^{1} y \,\, \overline{U}(x,y) dy
\label{flowraxi}
\end{equation}
and is known from the velocity profile.
From the profiles we measured, we can use (\ref{flowraxi}) as a measure of axi-symmetry. Indeed if the flow were axi-symmetric (\ref{flowraxi})
would give the same value at all stations.
An equivalent flow rate can be computed from the profiles obtained after the fractal orifice as if the flows were axi-symmetric.
In practice, such an equivalent flow rate will converge towards the actual volume flow rate as we get further away from the plate
and the flow reverts to an axi-symmetric shape.
There is another factor to take into account which is that the velocity we measure is $\overline{U}_{hwa}(y)$ not $\overline{U}(y)$.
So, large velocity fluctuations along the $y$ axis and recirculation effects near the wall will also affect our estimation of the flow rate.
However, $\overline{U}_{hwa}(y) \not\simeq \overline{U}(y)$ is also an effect of the fractal shapes and
calculating the volume flow rate from the profiles given in \fref{s1profilebymax} and
\fref{s2profilebyinlet} will still reveal the influence of the fractal orifices on the flow
and how they force or delay the return to axi-symmetric profiles.
\\[2ex]
In \fref{flowrate1} we plot ${\cal V}(x^*)/{\cal V}_{0}$, the guessed flow rate normalised by the inlet flow rate.
(a) corresponds to the mono-orifice type plates from from Set~1 and
(b) to the complex-orifice type plates from Set~2.
\\[2ex]
\begin{figure}[ht]
\includegraphics[width=7.5cm]{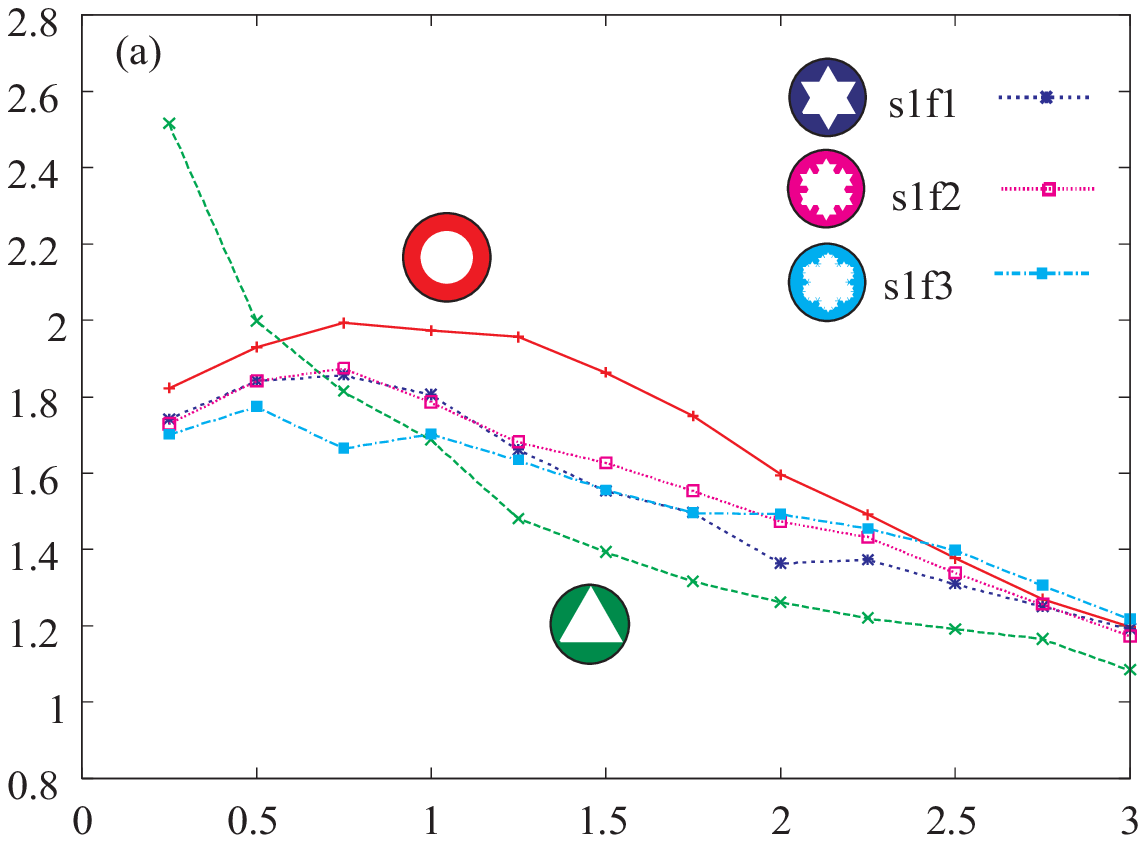}
\includegraphics[width=7.5cm]{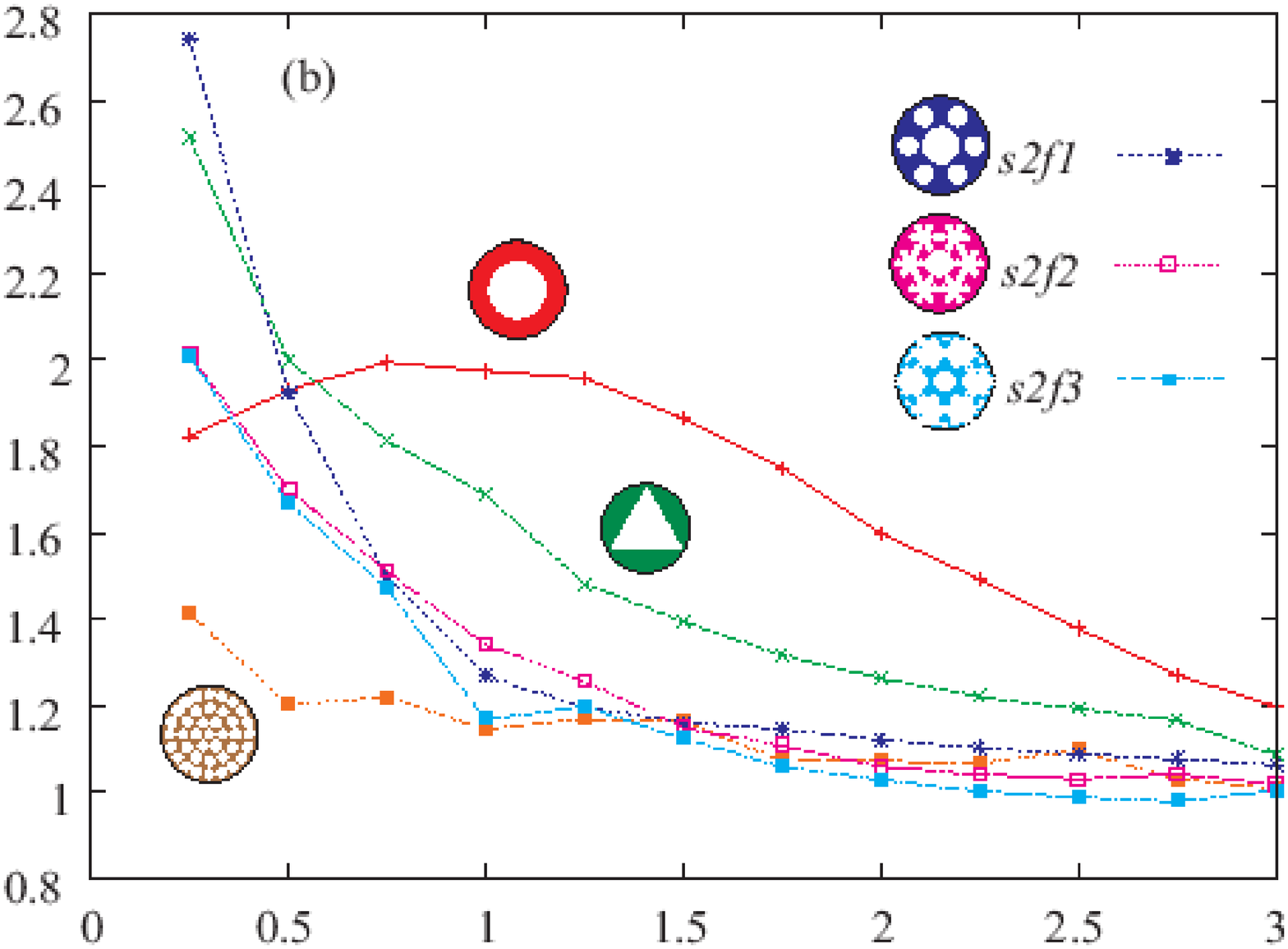}
\caption[ ]{Evolution of ${\cal V}/{\cal V}_{0}$ as a function of $x^*$ for the two types of plates: (a) Set~1, (b) Set~2.}
\label{flowrate1}
\end{figure}
%
As noticed before, the triangle orifice behaviour in \fref{flowrate1}a is at odds with the other orifice plates.
It is clear from \fref{flowrate1}a that its behaviour is closer to the behaviour for complex orifices though it is obviously a mono-orifice.
Omitting the triangle plate one can conclude that the guessed flow rate for the mono-orifices reflect the `vena contracta' effect.
The streamlines are deflected by the orifice resulting in increased velocities near the axis and the development of a re-circulation zone (with negative velocities)
near the wall.
The hotwire only measures the velocity modulus down to $U \simeq 3$~m~s$^{-1}$. So our guessed volume flow rate is clearly an overestimation in the
vena contacta region. From \fref{flowrate1} we can estimate the position of the vena contracta and of the maximum thickness of the recirculation zone.
This is around $x^*=1$ for the circular orifice and is closer to the plate for the fractal orifice around $x^*=0.75$ for \textit{s1f2} and \textit{s1f3}
and probably closer for \textit{s1f3} but we do not have the necessary precision to conclude for that orifice plate.
(The vena contracta effect also explains the difficulty encountered in \citep{Nicolleau-et-al-JOT-2011} when trying to apply directly
\cite{Mazellier-Vassilicos-2010}'s evolution law for the maximum of $u'$.)
\\[2ex]
Further down all the profiles converge to the same asymptotic curve after $x^*=2.25$.
But the trend is still to a decrease and we can conclude that the flow has not yet fully reverted to axi-symmetry at $x^*=3$.
All the fractal orifices profiles for \textit{s1f1}, \textit{s1f2} and \textit{s1f3} are similar. So that we can conclude that in terms of return to axi-symmetry
the first fractal level \textit{s1f1} is already optimum.
\\[2ex]
\Fref{flowrate1}b shows the evolution of the guessed flow rate for the perforated type plates
from Set~2. They all show the same trend near the plate that is an over-prediction of the flow rate and a monotonic decrease as $x^*$ increases.
This is also the trend followed by the triangular-shaped orifice. This is actually due to the interaction of the flow coming from the plate and the boundary
layer near the wall. Indeed the common feature between all these plates including the triangle orifice is the smallest distance from the wall to the flow area.
To characterise this effect we define a new non dimensional number $\delta_g^*$ as
\begin{equation}
\delta_g^* = \delta_g/D
\end{equation}
where $\delta_g$ is the smallest distance from the flow area to the wall. Values of $\delta_g^*$ are reported in \tref{tabpara}.
The complex-orifices and triangular shaped orifice have all $\delta_g^* < 0.06$, whereas the other orifice-like plates have all $\delta_g^*>0.11$.

\section{Velocity fluctuation RMS profiles}
\label{rmsprof}

\subsection{Set 1: RMS profiles}

The rms profiles behind the plates in Set~1 are shown in \fref{s1rms}.
\begin{figure}[ht!]
\includegraphics[width=16cm]{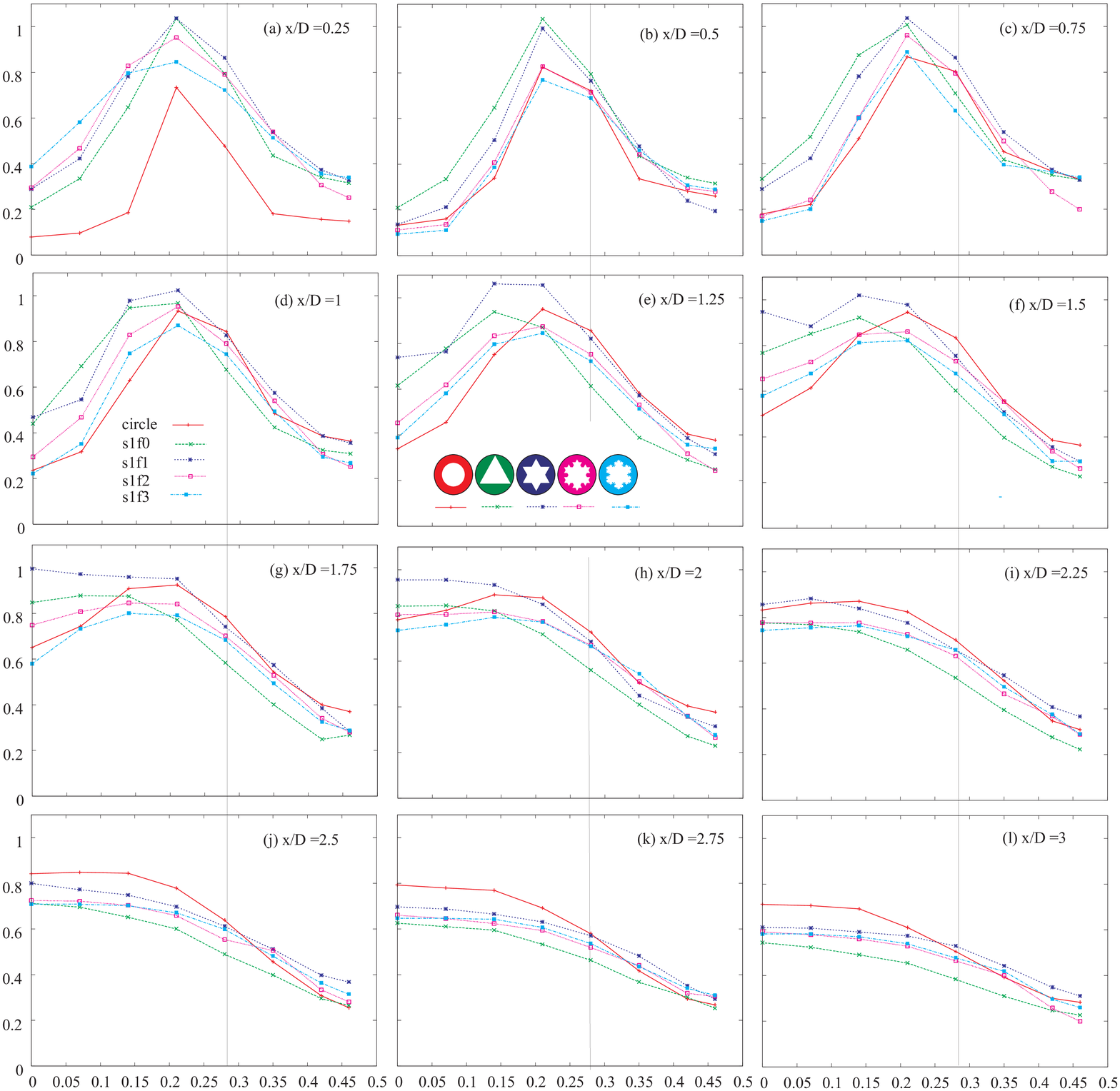}
\caption[]{Set~1: $u'/U_0$ (vertical axis) as functions of $y^*$ (horizontal axis) at different station $x^*$.
The vertical lines crossing the plots indicate the position of the circular orifice's edge.}
\label{s1rms}
\end{figure}
%
The position of the circular orifice's edge is indicated by a light vertical line. The fractal perimeters have no much effect on the rms outside the area of the circular orifice except for the triangular orifice which apex is clearly outside the area delimitated by the circular orifice (for Set~1 $\delta_g^*$ varies from 0.05 for the triangular orifice to 0.21 for the circular orifice) and has an influence on the boundary layer developing from the pipe's wall.
\\[2ex]
Most of the effect of the fractal edges is therefore concentrated within the area delimitated by the circular shape. Within that area, there is little effect of the fractal perimeter on
the position of the maximum for $u'$.
Near the plate ($x^*=0.25$) the
peak of $u'$ is observed at the same position for all plates, that is around $y^*=0.2 \pm 0.05$.
For the circular orifice,
for as long as there is a peak for $u'$ it stays
at $y^*=0.2$.
\\[2ex]
Close to the plate up to $x^{*} \simeq 1D$, the fractal orifices give higher turbulence fluctuations than the classical circular orifice and the higher the fractal iteration the lower $u'$.
For $x^{*} > 1D$ the triangle's profile departs from the general trend followed by the fractal perimeter, its overall profile is now significantly influenced by the portion outside the area delimitated by the circular shape.
For the other plates
\textit{s1f1}, \textit{s1f2}, \textit{s1f3}, the higher the fractal iteration the lower $u'$.
This is observed at any distance from the orifice.
\\[2ex]
Further than $x^{*}=2.25$ from the plate, the fractal orifices yield lower turbulence
fluctuations than the circular orifice within the circular delimitated area.
They also generate flatter profiles indicating a faster flow recovery.
As expected the further from the plate the less differences between the different fractal orifices. There is little differences between \textit{s1f2}, \textit{s1f3} as close to the plate
as $x^*=2$.

\subsection{RMS profiles Set~2}

\Fref{s2rms} shows the rms profiles behind the different complex-orifices from Set~2. For the sake of comparison the profiles from the classical perforated plate
are added to the plots and the position of the circular orifice's edge is indicated by a light vertical line.

\begin{figure}[ht!]
\includegraphics[width=16cm]{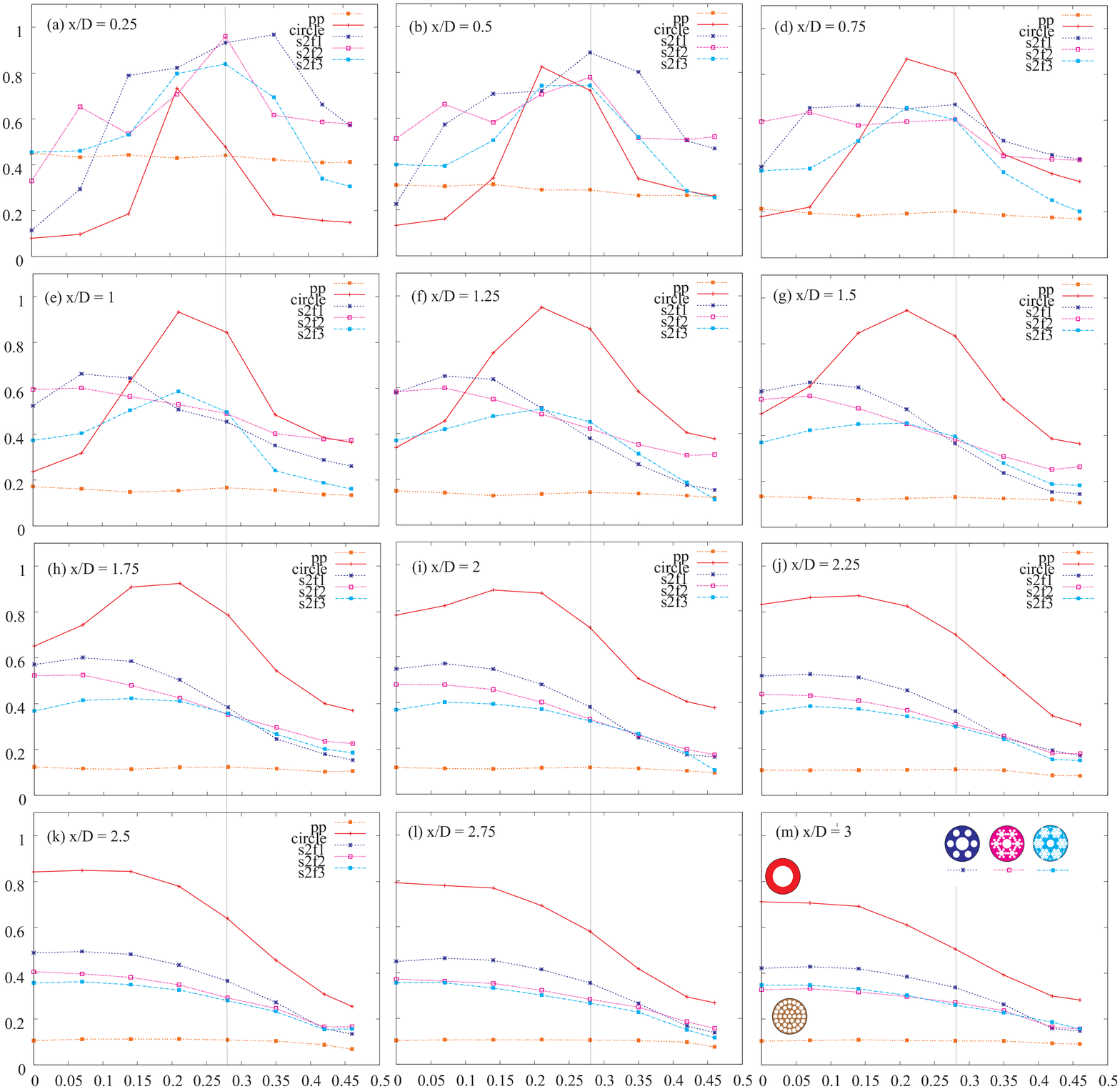}
\caption[]{Set~2: $u'/U_0$ (vertical axis) as functions of $y^*$ (horizontal axis) at different stations $x^*$}
\label{s2rms}
\end{figure}
%
The effect of the complex-orifices is clear from the onset: just after the orifice, there is not a single-peaked profile as for the circular orifice but the profiles for the complex-orifices are higher and flatter. They are also higher than that of the perforated plate except at the pipe centre for the lower order fractal orifices.
\\[2ex]
Far from the plate position, the profiles seem to converge towards the flat level profile obtained with the perforated plate.
There is a clear effect of the fractal iteration level on that convergence: the higher the level, the closer to the perforated plate results.
After $x^*=2.25$ there are little differences between \textit{s2f2} and \textit{s2f3}.
\\[2ex]
Similarly to Set~1, the fractal orifices first give turbulence levels higher than those of the circular orifice.
Then the complex-orifices' turbulence fluctuation decreases rapidly while that of the circular orifice increases before reaching a smoother-shaped profile.
As can be seen in \fref{s2rms}a, by contrast to Set~1 the complex orifices are acting outside the area delimitated by the reference circular orifice. So eventually their profile get lower than that of the circular orifice even outside the delimitated area.
After $x^{*}=1.75$ the complex-orifices' profiles are always smaller than the circular orifice's profile and much closer to the perforated plate profile.
\\[2ex]
To quantify the production of turbulence energy $u'^2$ relative to each orifice
we can define the non-dimensional parameter $\alpha$ as follows:
\begin{equation}
\alpha_i(x^*) = \max_{y^*}{u'_i(x^*) \over u'_p(x^*)}
\end{equation}
where $i$ designates a particular orifice and $p$ stands for the classical perforated orifice that we take as our reference.
$\alpha$ is defined for each orifice as a function of $x^*$. It measures the production of velocity fluctuation relative to the classical perforated orifice.
The values of $\alpha$ are reported in \fref{alpha} for Set~1 and Set~2.
\begin{figure}[ht!]
\includegraphics[width=7.7cm]{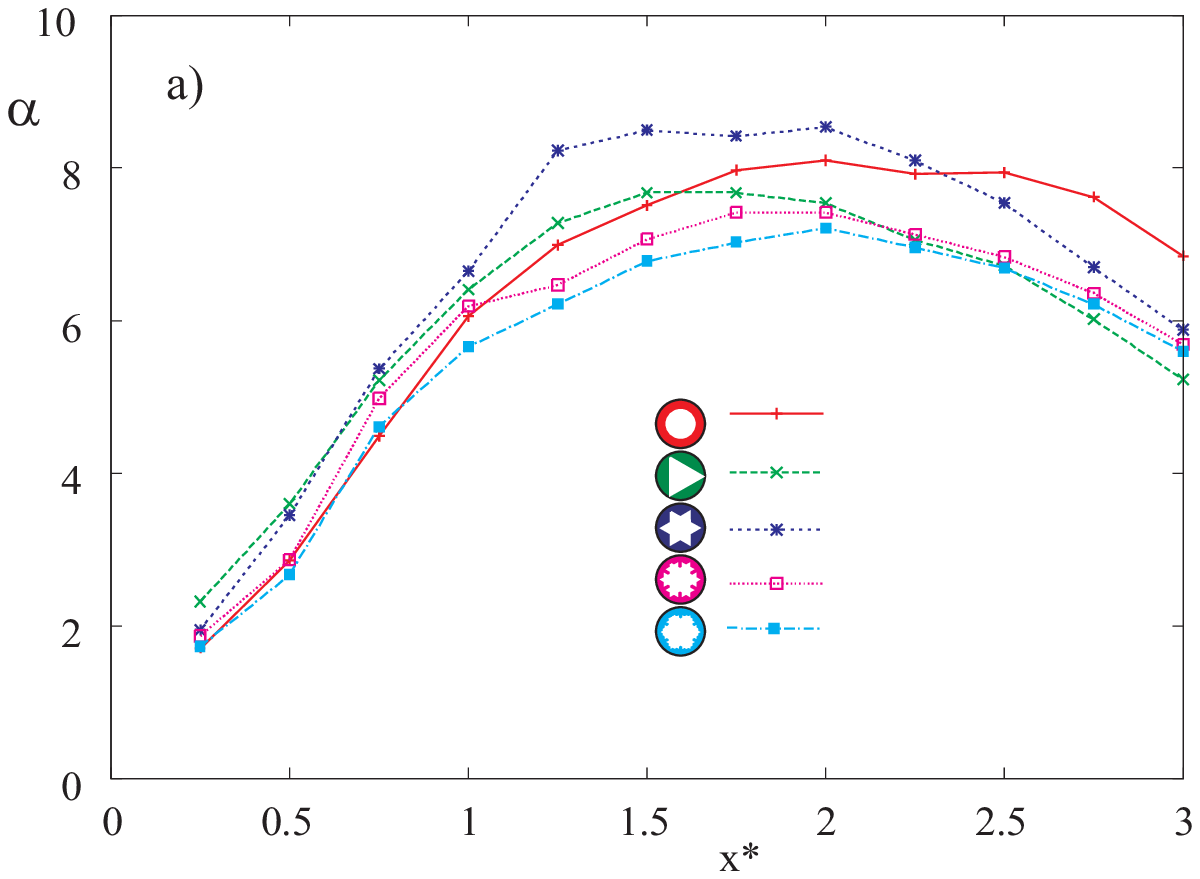}
\includegraphics[width=7.7cm]{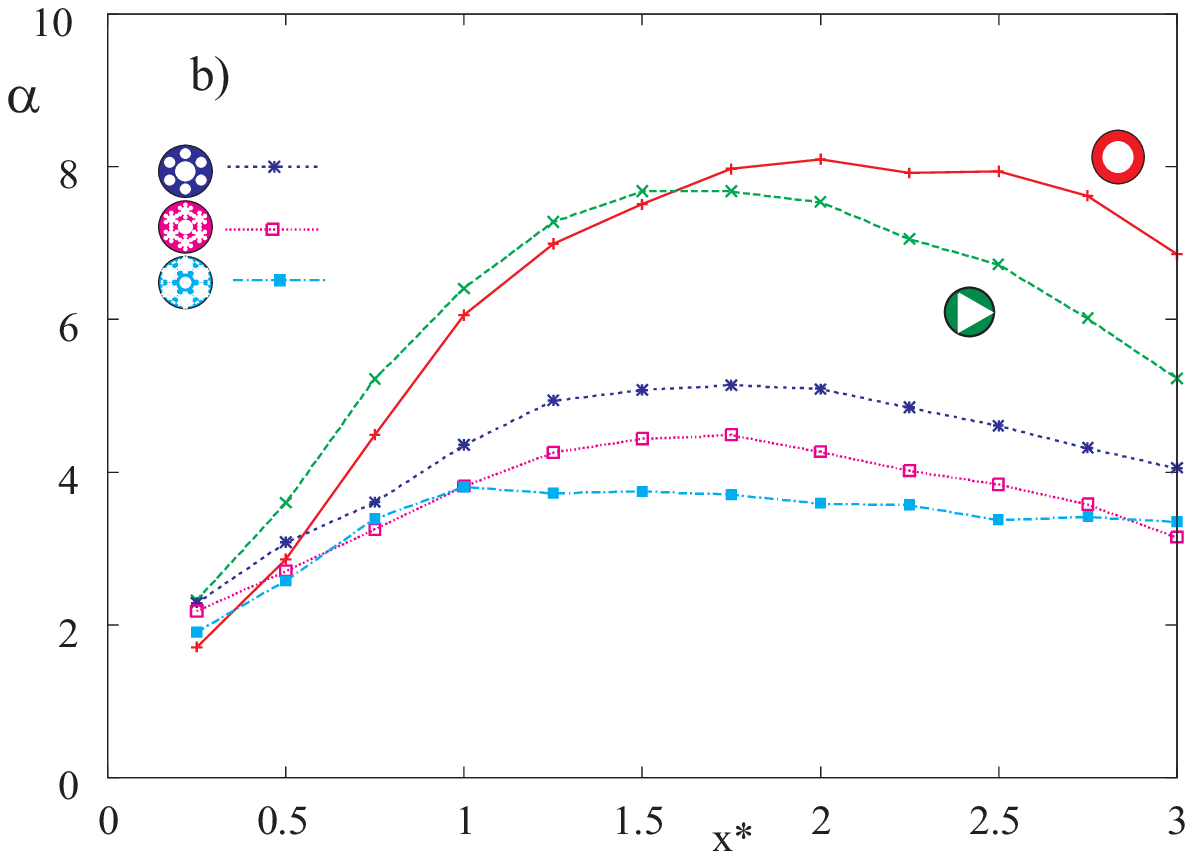}
\caption[]{$\alpha$ as a function of $x^*$ for the different orifices in a) Set~1, b) Set~2}
\label{alpha}
\end{figure}
\\
All the orifices show the same trend for $\alpha$, it first increases and then decreases showing a maximun in the range $1 \le x^* \le 2$.
There is a clear fractal iteration effect: the higher the iteration the lower $\alpha$.
\textit{s2f3} as the highest iteration looks as an interesting limiting case: the decrease of $\alpha$ with $x^*$ is very small and it seems that the flow after \textit{s2f3} will retain a much higher turbulence level than the reference perforated plate for a very long distance.
\\[2ex]
The different iterations in both sets correspond to different scale ratios (see \tref{tabpara}).
Furthermore, in Set~2, different connexity parameters are linked to different scale ratios. These two quantities basically tell us the range of scales forced at the orifice (scale ratio) and how many of them there are (how many holes is the connexity parameter). So decreasing the connexity, (and scale ratio) is like forcing on a larger range of scales and a less confined physical space and indeed that flattens the rms profiles and decreases the maximum values reflected by $\alpha$.


\section{Conclusion}

It was shown in \citep{El-Azm-et-al-2010,Nicolleau-et-al-JOT-2011}
that fractal orifices yield lower pressure drops than the popular circular orifice; a very important property for flowmetering applications.
That initial study needed to be completed by an assessment of how disruptive to the flow the fractal-orifices are when compared to the two benchmarks
that are the circular orifice and perforated plate. (This latter was chosen for its standard flow mixing properties as it is not relevant of course for flowmetering.) In particular, a question one may ask is: how long it takes for the fractal-generated flow to return to axi-symmetry and
smooth out the high turbulence levels it may generate?

This paper compares the merits of different
orifice plates for the two reference classes: mono-orifice and complex-orifice,
for a rapid return to axi-symmetric flow.
The guessed flow rate is introduced as an objective measure of how disruptive is the
orifice to the flow in view of flowmetering techniques.

The results show that for the mono-orifice an important feature to consider is the interaction of the object with the re-circulation zone from the wall.
The parameter $\delta_g^*$ is introduced to measure the smallest gap between the flow area and the wall.
For the mono-orifice with $\delta_g^* > 0.05$ the return to axi-symmetry is better than that of the circular orifice but there is no significant
effect of the fractal iteration. (At least for the Reynolds number used here.)

For the complex-orifice the return to axi-symmetry is always better than that of the circular orifice. There is a clear effect of the fractal iteration near the plate,
though as for the mono-orifice, for that Reynolds number there is an optimum iteration ($s2f2$) above which there is no significant effect of the iteration.

In terms of turbulence intensity as measured by the factor $\alpha$ the fractal-orifices are always better than the circular orifice after $\sim 2.5$ diameter.
The complex-orifices seem particularly good for that purpose.
\\[2ex]
Overall, the results confirm the potential of fractal orifices as flowmeters. It was previously known that they are more efficient in terms of pressure drop. The present
study indicates that they are also more efficient in re-generating an axi-symmetric flow with lower level of velocity fluctuations. This is also very important for
applications to flowmetering where a standard flow after the orifice is required for robust accurate measurements.
\\[2ex]
For the Reynolds number used here, there is an optimum iteration of the fractal level above which no improvement to the
flowmeter is to be expected. So we expect in future works to be able to define the range of useful iterations for a given range of Reynolds numbers.
The main parameters proposed for the characterisation of the fractal orifice are the
connexity parameter, the symmetry angle and gap to the wall $\delta^*_g$. But more work with more geometries needs to be done to
investigate all these parameters.
\\[2ex]
There is also a need for validation of more complex subgrid models dealing in particular
with rough surfaces \citep{Laizet-Vassilicos-2009,Anderson-Meneveau-2011,Lopez-et-al-2011,Melick-Geurts-2012}
and fractal-forced flows could provide a systematic way to generate data for such validations.
In particular it is easy to see that such flows pose a real challenge to grid-dependent method as Detached Eddy Simulation~\citep{Zheng-et-al-2009-Gdansk}.
%


\ack
The authors would like to thanks the Leverhulme Trust for its support through grant F/00~118/AZ.

\section*{References}
\begin{harvard}
\harvarditem{Abou-El-Azm et~al.}{2010}{El-Azm-et-al-2010}
Abou-El-Azm A, Chong C, Nicolleau F \harvardand\ Beck S 2010 {\em Experimental
  Thermal and Fluid Sc.} {\bf 34},~104--111.
doi:10.1016/j.expthermflusci.2009.09.008.

\harvarditem{Anderson \harvardand\ Meneveau}{2011}{Anderson-Meneveau-2011}
Anderson W \harvardand\ Meneveau C 2011 {\em J. Fluid Mech.} {\bf
  679},~288–314.
doi:10.1017/jfm.2011.137.


\harvarditem{Farge et~al.}{1993}{Farge-al93}
Farge M, Hunt J \harvardand\ Vassilicos J 1993 {\em Wavelets, fractals and
  Fourier transforms} Clarendon Press, Oxford.

\harvarditem{Laizet \harvardand\ Vassilicos}{2009}{Laizet-Vassilicos-2009}
Laizet S \harvardand\ Vassilicos J~C 2009 {\em Journal of Multiscale
  Modelling} {\bf 1}(1),~177--196.

\harvarditem{Lopez Penha et~al.}{2011}{Lopez-et-al-2011}
Lopez Penha D~J, Geurts B~J, Stolz S \harvardand\ Nordlund M 2011
{\em Computers and Fluids} {\bf 51},~157-173.

\harvarditem{Manshoor et~al.}{2011}{Manshoor-et-al-2011}
Manshoor B, Nicolleau F \harvardand\ Beck S 2011 {\em Flow Measurement and
  Instrumentation} {\bf 22}(3),~208--214.
doi: 10.1016/j.flowmeasinst.2011.02.003.

\harvarditem{Mazellier \harvardand\
  Vassilicos}{2010}{Mazellier-Vassilicos-2010}
Mazellier N \harvardand\ Vassilicos JC 2010 {\em Phys. Fluids} {\bf
  22}(7),~07510.
doi:10.1063/1.3453708.

\harvarditem{van~Melick \harvardand\ Geurts}{2012}{Melick-Geurts-2012}
van~Melick P.A.J \harvardand\ Geurts B.J. 2012
{\em Fluid Dyn. Res.}

\harvarditem{Meneveau}{1991}{Meneveau-1991}
Meneveau C 1991 {\em J. Fluid Mech.} {\bf 224},~429--484.

\harvarditem{Meneveau \harvardand\
  Sreenivasan}{1990}{Meneveau-Sreeninvasan-1990}
Meneveau C \harvardand\ Sreenivasan K~R 1990 {\em Physical Review A} {\bf
  41}(4),~2246--2248.

\harvarditem{Michelitsch et~al.}{2009}{Michelitsch-et-al-2009}
Michelitsch T, Maugin G, Nicolleau F, Nowakowski A \harvardand\ Derogar S 2009
  {\em Phys. Rev. E} {\bf 80},~011135.
doi:10.1103/PhysRevE.80.011135.

\harvarditem{Nastase et~al.}{2011}{Ilinca-et-al-2011}
Nastase I, Meslem A \harvardand\ Hassan M~E 2011 {\em Fluid Dyn. Res.} {\bf
  43},~065502.
doi:10.1088/0169-5983/43/6/065502.

\harvarditem{Nicolleau}{1996}{Nicolleau-1996}
Nicolleau F 1996 {\em Phys. Fluids} {\bf 8}(10),~2661--2670.

\harvarditem{Nicolleau et~al.}{2011}{Nicolleau-et-al-JOT-2011}
Nicolleau F, Salim S \harvardand\ Nowakowski A 2011 {\em Journal of
  Turbulence} {\bf 12}(44),~1--20.
doi: 10.1080/14685248.2011.637046.

\harvarditem{Nicolleau \harvardand\ Vassilicos}{1999}{Nicolleau-Vassilicos99}
Nicolleau F \harvardand\ Vassilicos J 1999 {\em Phil. Trans. R. Soc. Lond. A}
  {\bf 357}(1760),~2439--2457.
doi: 10.1098/rsta.1999.0441.

\harvarditem{Nowakowski et~al.}{2011}{Nowakowski-et-al-AIAA-2011}
Nowakowski A, Nicolleau F \harvardand\ Salim S 2011 {\em in} `6th AIAA
  Theoretical Fluid Mechanics Conference' Hawaii Convention Center Honolulu,
  Hawaii, USA p.~(ID: 1023232).

\harvarditem{Seoud \harvardand\ Vassilicos}{2007}{Seoud-Vassilicos-2007}
Seoud R~E \harvardand\ Vassilicos J~C 2007 {\em Phys. Fluids} {\bf
  19}(10),~105108.

\harvarditem{Sreenivasan et~al.}{1989}{Sreenivasan-al-1989}
Sreenivasan K~R, Ramshankar R \harvardand\ Meneveau C 1989 {\em Proc. R. Soc.
  Lond. A} {\bf 421},~79--108.

\harvarditem{Valente \harvardand\ Vassilicos}{2012}{Valente-Vassilicos-2012}
Valente P~C \harvardand\ Vassilicos J~C 2012 {\em Phys. Lett. A} {\bf
  376},~510--514.

\harvarditem{Zheng et~al.}{2010}{Zheng-et-al-2009-Gdansk}
Zheng H, Nicolleau F \harvardand\ Qin N 2010 {\em in} S~Peng, P~Doerffer
  \harvardand\ W~Haase, eds, `Notes on Numerical Fluid Mechanics: 3rd Symposium
  on Hybrid RANS-LES Methods' Vol. 111 Gdansk, Poland Gdansk, Poland
  pp.~157--165.
doi: 10.1007/978-3-642-14168-3-13.

\end{harvard}

\end{document}

  Lond. A} {\bf 421},~79--108.

\harvarditem{Valente \harvardand\ Vassilicos}{2012}{Valente-Vassilicos-2012}
Valente P~C \harvardand\ Vassilicos J~C 2012 {\em Phys. Lett. A} {\bf
  376},~510--514.

\harvarditem{Zheng et~al.}{2010}{Zheng-et-al-2009-Gdansk}
Zheng H, Nicolleau F \harvardand\ Qin N 2010 {\em in} S~Peng, P~Doerffer
  \harvardand\ W~Haase, eds, `Notes on Numerical Fluid Mechanics: 3rd Symposium
  on Hybrid RANS-LES Methods' Vol. 111 Gdansk, Poland Gdansk, Poland
  pp.~157--165.
doi: 10.1007/978-3-642-14168-3-13.

\end{thebibliography}


\end{document}